%% file: 0_main.tex
\documentclass[conference,10pt]{IEEEtran}
\IEEEoverridecommandlockouts
\usepackage{cite}
\usepackage{amsmath,amssymb,amsfonts}
\usepackage{algorithmic}
\usepackage{graphicx,physics,array}
\usepackage{textcomp}
\usepackage{xcolor,soul}
\usepackage{multirow}
\usepackage{hyperref}
\usepackage{adjustbox}
\usepackage{dsfont}

\usepackage{tikz}
\newcommand*\circled[1]{\tikz[baseline=(char.base)]{
            \node[shape=circle,fill,inner sep=0.5pt] (char) {\textcolor{white}{#1}};}}

\def\BibTeX{{\rm B\kern-.05em{\sc i\kern-.025em b}\kern-.08em
    T\kern-.1667em\lower.7ex\hbox{E}\kern-.125emX}}

\begin{document}

\title{Lattice Surgery for Dummies
}

\author{\IEEEauthorblockN{Avimita Chatterjee}
\IEEEauthorblockA{\textit{Department of Computer Science \& Engineering} \\
\textit{Pennsylvania State University}\\
PA, USA\\
amc8313@psu.edu}
\and
\IEEEauthorblockN{Subrata Das}
\IEEEauthorblockA{\textit{School of EECS} \\
\textit{Pennsylvania State University}\\
PA, USA\\
sjd6366@psu.edu}
\and
\IEEEauthorblockN{Swaroop Ghosh}
\IEEEauthorblockA{\textit{School of EECS} \\
\textit{Pennsylvania State University}\\
PA, USA\\
szg212@psu.edu}
}

\maketitle

\begin{abstract}
Quantum error correction (QEC) plays a crucial role in correcting noise and paving the way for fault-tolerant quantum computing. This field has seen significant advancements, with new quantum error correction codes emerging regularly to address errors effectively. Among these, topological codes, particularly surface codes, stand out for their low error thresholds and feasibility for implementation in large-scale quantum computers. However, these codes are restricted to encoding a single qubit. Lattice surgery is crucial for enabling interactions among multiple encoded qubits or between the lattices of a surface code, ensuring that its sophisticated error-correcting features are maintained without significantly increasing the operational overhead. Lattice surgery is pivotal for scaling QECCs across more extensive quantum systems. 
Despite its critical importance, comprehending lattice surgery is challenging due to its inherent complexity, demanding a deep understanding of intricate quantum physics and mathematical concepts. This paper endeavors to demystify lattice surgery, making it accessible to those without a profound background in quantum physics or mathematics. This work explores surface codes, introduces the basics of lattice surgery, and demonstrates its application in building quantum gates and emulating multi-qubit circuits.
\end{abstract}


\input{1_introduction}
\input{2_theory}
\input{3_lattice_surgery}
\input{4_gate_operations}
\input{5_multi_qubit}
\input{7_conclusion}

\section*{Acknowledgment}

The work is supported in parts by the National Science Foundation (NSF) (CNS-1722557, CCF-1718474, OIA-2040667, DGE-1723687, and DGE-1821766) and gifts from Intel.

\bibliographystyle{IEEEtran}
\bibliography{references}

\end{document}

%% file: 1_introduction.tex
\section{Introduction} \label{intro}

Quantum computing leverages quantum mechanics principles to perform tasks unachievable by classical computing, with applications spanning molecule simulation for drug development, financial modeling improvements, machine learning advancements, optimization task enhancements, and supply chain management transformations \cite{reiher2017elucidating, orus2019quantum, schuld2015introduction, gachnang2022quantum, ajagekar2019quantum}. Yet, the path to commercializing these innovations faces hurdles, including issues with qubit stability and quantum noise \cite{clerk2010introduction, mouloudakis2021entanglement}. Quantum Error Correction Codes (QECCs) play a pivotal role in realizing fault-tolerant quantum computing amidst inherent qubit noise \cite{lo1998introduction, devitt2013quantum}. Unlike conventional error correction approaches \cite{hamming1950error}, quantum error correction encounters specific obstacles due to the no-cloning theorem \cite{wootters2009no} and the phenomenon of wave-function collapse during qubit measurement \cite{von2018mathematical}. This ongoing research has yielded a variety of quantum codes, including five-qubit, Bacon-Shor, topological, surface, color, and heavy-hexagon codes \cite{sundaresan2022matching, abobeih2022fault, bacon2006operator, kitaev1997quantum, krinner2022realizing, bombin2006topological}, each contributing to the advancement towards fault-tolerant quantum computation.

\subsection{Motivation}

All QECCs are designed to encode a single qubit. When faced with a circuit involving multiple qubits that require QECC application, enabling interactions among these encoded qubits becomes necessary. QECCs typically employ transversal gates for this purpose which offer scalability but increase gate overhead significantly diminishing the QECCs' error-correcting capabilities. This challenge is addressed by lattice surgery \cite{horsman2012surface}, a technique that enables interactions between multiple encoded qubits without compromising their error-correcting properties while keeping the increase in overhead manageable. Lattice surgery is essential to achieve fault-tolerant quantum computing, as it ensures QECCs can scale up to larger systems. Predominantly associated with surface codes, among the most prominent QECCs due to their low error thresholds and practicality for large-scale implementation, lattice surgery's principles also lay the groundwork for advancing other QECCs \cite{landahl2014quantum}. Examples of such work include the qubit lattice surgery \cite{cowtan2022qudit} and lattice surgery for color-codes \cite{landahl2014quantum}. Understanding lattice surgery is thus crucial for developing fault-tolerant quantum computing. 

\subsection{Contribution}

While grasping lattice surgery is vital, it presents significant challenges due to the sophisticated mathematical concepts of quantum physics, especially concerning surface codes and qubit interactions. The research papers on lattice surgery \cite{ erhard2021entangling, fowler2018low, litinski2019game, herr2018lattice, chamberland2022circuit, chamberland2022universal, herr2017optimization, de2020zx, vuillot2019code} often assume a deep understanding of such concepts, making it difficult for newcomers to follow. This work aims to offer a clear and approachable overview of lattice surgery's key principles, tailored for researchers with limited experience in quantum physics or its mathematical underpinnings. This work provides a comprehensive overview of surface codes before delving into the fundamental principles of lattice surgery. It begins with the basics of lattice surgery, progresses through the construction of quantum gates using this technique, and ultimately demonstrates how lattice surgery can be used to emulate multi-qubit circuits. Prior familiarity with surface codes is not required for readers of this review. However, a basic acquaintance with quantum circuit symbols, as outlined in \cite{nielsen2001quantum}—including fundamental measurement techniques, the controlled-NOT (CNOT) gate, and the Hadamard (H) gate—is presumed.

\subsection{Paper Structure}

The paper begins by laying a theoretical foundation on QECCs, surface codes, and transversal gates in Section \ref{theory}. It then outlines the objectives of lattice surgery and its classifications in Section \ref{lattice_surgery} and the range of operations achievable through lattice surgery in Section \ref{gate_operations}. The application of QECCs to circuits involving multiple qubits, facilitated by lattice surgery, is discussed in Section \ref{multi_qubit}. 
Conclusions are drawn in Section \ref{conclusion}.

%% file: 2_theory.tex
\section{Theoretical Background} \label{theory}

\begin{figure*}
    \centering
    \includegraphics[width=1\linewidth]{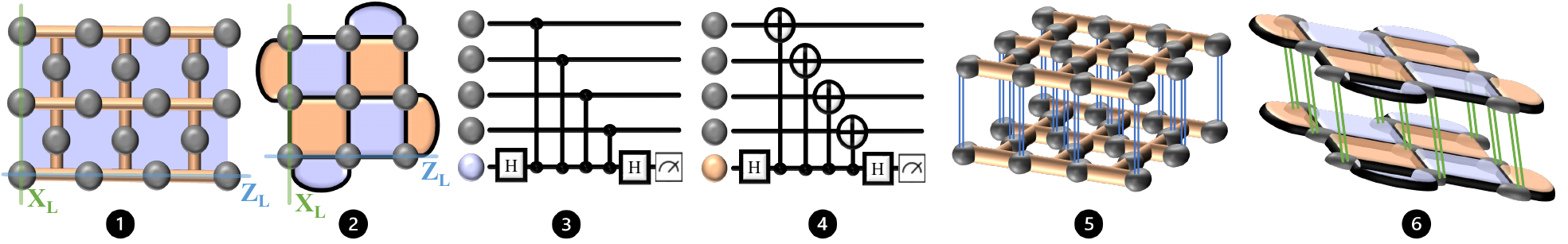}
    \caption{\textbf{Surface Codes.} 
    \raisebox{.5pt}{\textcircled{\raisebox{-.9pt} {\textbf{1}}}}: Structure of an unrotated surface code.
    \raisebox{.5pt}{\textcircled{\raisebox{-.9pt} {\textbf{2}}}}: Structure of an rotated surface code.
    \raisebox{.5pt}{\textcircled{\raisebox{-.9pt} {\textbf{3}}}}: Circuit representation of a Z-stabilizer acting on four qubits and projecting its syndrome onto an ancilla qubit which is later measured.
    \raisebox{.5pt}{\textcircled{\raisebox{-.9pt} {\textbf{4}}}}: Circuit representation of an X-stabilizer acting on four qubits and projecting its syndrome onto an ancilla qubit which is later measured.
    \raisebox{.5pt}{\textcircled{\raisebox{-.9pt} {\textbf{5}}}}: Visualization of transversal 2-qubit gates between two unrotated surfaces.
    \raisebox{.5pt}{\textcircled{\raisebox{-.9pt} {\textbf{6}}}}: Illustration of transversal 2-qubit gates between two rotated surfaces.
    }
    \label{fig:surface_code}
\end{figure*}

\subsection{A Brief Overview of QECCs}


Quantum Error-Correcting Codes (QECCs) are designed to protect quantum information from errors due to decoherence and other quantum noise, thereby enabling reliable quantum computation \cite{chatterjee2023quantum, devitt2013quantum}. Understanding the interaction between physical and logical qubits, the encoding process, the types of errors encountered, and the use of stabilizers and ancilla qubits is fundamental in implementing QECCs effectively.

In quantum computing, physical qubits are the basic units of quantum information, analogous to bits in classical computing. These qubits can exist in states represented by superposition, allowing them to hold a combination of 0 and 1 simultaneously, and can become entangled with each other, a property that underpins the power of quantum computing. Logical qubits, however, are formed by encoding quantum information across multiple physical qubits. This encoding uses redundancy to enhance fault tolerance against noise and errors \cite{devitt2013quantum, chiaverini2004realization}. Logical qubits are thus more stable constructs designed to implement the robust storage and manipulation of information in a noisy quantum environment. The encoding of quantum information into logical qubits involves distributing the state of a single qubit across a group of physical qubits. This distribution is managed through specific quantum gates that entangle the qubits in a way that any error affecting a single physical qubit can be detected and corrected without collapsing the overall quantum state. The robustness of a QECC is largely defined by how effectively it encodes and preserves the state of logical qubits despite errors in physical qubits.

Quantum computations contend primarily with two types of errors: (i) Bit-flip errors that flip the state of a qubit from $\ket{0}$ to $\ket{1}$ or vice versa. This type of error is analogous to the classical bit flip and can disrupt computations by altering the basic state of a qubit and (ii) Phase-flip errors that alter the relative phase between the basis states of a qubit. In a superposed state, this error can change the relative weighting of $\ket{0}$ and $\ket{1}$, leading to a loss of coherence in the quantum information. Stabilizers are operators used in QECCs to check for errors without measuring the qubits directly, which would otherwise collapse their quantum state. There are typically two types of stabilizers used. Z-type stabilizers detect bit-flip errors by checking for unexpected changes in the qubit states. X-type stabilizers are used to detect phase-flip errors by observing changes in the phase relationships between qubits.

Ancilla qubits are additional qubits included in a QECC to aid in the error detection process \cite{chiaverini2004realization}. They do not hold quantum information themselves but are used to interact with logical qubits to extract error syndromes—a form of indirect measurement that indicates whether and where an error has occurred. During syndrome measurement, ancilla qubits are entangled with logical qubits and then measured. The outcome of these measurements is used to diagnose errors based on the known properties of the stabilizers. Through syndrome measurement, the QECC can determine not just the presence of an error but also its type and possible location (using suitable decoding algorithm), enabling targeted error correction that restores the integrity of the quantum information without needing to observe the logical qubits directly. This process underscores the sophisticated interplay between encoding, error detection, decoding and error correction that QECCs use to maintain the fidelity of quantum computations amidst the inherent fragility of quantum states.

\subsection{Surface Codes}


Surface codes are a significant advancement in the field of quantum error correction, distinguished by their two-dimensional lattice structure that contributes to their high error tolerance \cite{dennis2002topological}. Originating from the concept of toric codes, surface codes are a type of topological quantum error-correcting code that effectively utilizes the arrangement of qubits on a grid to safeguard quantum information against common quantum errors, specifically bit-flip and phase-flip errors \cite{krinner2022realizing, dennis2002topological}.

These codes employ a unique strategy involving ancilla qubits to monitor the integrity of the quantum state. These ancilla qubits are integral to performing X and Z stabilizer checks across the lattice, allowing for the detection of the two primary error types without disrupting the quantum state of the data qubits. The effectiveness of these stabilizers in diagnosing errors is a key feature of surface codes, leveraging Pauli operations to probe groups of qubits and thereby determine the presence and locations of errors.

In the domain of quantum computing, the n-distance surface code is depicted using an \(n \times n\) lattice configuration, where each position, or `blob', represents a qubit that is crucial to maintain the logical state of the system. The structure of this lattice ensures that every qubit is monitored by both X and Z stabilizers \cite{fowler2012surface}. These stabilizers are essential for identifying phase and bit-flip errors respectively, utilizing combinations of Pauli operators (X or Z) that act on specific subsets of qubits within the lattice. Should an error occur, it alters the outcome of the stabilizer checks linked to the affected qubit. By conducting these stabilizer checks, it becomes feasible to pinpoint the timing and location of any errors using a decoding algorithm \cite{kolmogorov2009blossom}. Once the presence and position of errors are confirmed, they can be rectified by applying quantum gates to flip the affected qubits back to their intended states, thus restoring the integrity of the quantum information.

The architecture of surface codes can be implemented in two variations: unrotated and rotated, each with a distinct lattice structure that affects their error-correction capabilities. The unrotated version features a square lattice where data qubits reside on the edges, and each square (or plaquette) links to a Z-stabilizer, with X-stabilizers located at the vertices connecting the squares \cite{fowler2012surface}. Figure \ref{fig:surface_code} \raisebox{.5pt}{\textcircled{\raisebox{-.9pt} {\textbf{1}}}} shows an unrotated surface code, where qubits represented by grey blobs are influenced by Z-stabilizers depicted as purple squares and X-stabilizers illustrated with orange lines.

Conversely, the rotated version uses a tilted lattice that alternates X and Z stabilizers in a checkerboard pattern, impacting the qubits at each box's vertices \cite{tomita2014low}. The stabilizers engage with the qubits located at the vertices of each specific box. Figure \ref{fig:surface_code} \raisebox{.5pt}{\textcircled{\raisebox{-.9pt} {\textbf{2}}}} displays a rotated surface code where purple surfaces represent Z-stabilizers and orange surfaces indicate X-stabilizers. Together, these stabilizers interact with all the qubits depicted as grey blobs, which form the logical state. This rotated arrangement often provides a slightly higher error threshold and is simpler to implement, making it more advantageous for robust quantum error correction over long distances.

When examining the functionality of these stabilizers in a quantum circuit, there are typically two main methodologies employed to demonstrate their interaction with the qubits. Figure \ref{fig:surface_code} \raisebox{.5pt}{\textcircled{\raisebox{-.9pt} {\textbf{3}}}} illustrates the creation of a Z-stabilizer using a combination of CZ and Hadamard gates. This method enhances the Z-stabilizer's capability to effectively detect bit-flip errors through a sophisticated manipulation of the quantum state. The configuration affects an ancillary qubit, which is visually depicted as a purple blob within the diagrams. After the interaction facilitated by the CZ and Hadamard gates, the state of this ancillary qubit is measured to provide a syndrome measurement, denoted as S\textsubscript{z}. This measurement is crucial as it reflects any errors detected across the qubits that the Z-stabilizer governs, showcasing the effectiveness of this gate combination in maintaining the integrity of the quantum state.

Figure \ref{fig:surface_code} \raisebox{.5pt}{\textcircled{\raisebox{-.9pt} {\textbf{4}}}} demonstrates the construction of an X-stabilizer through the use of CNOT and Hadamard gates. This arrangement is designed to detect phase-flip errors, employing these gates to adjust the quantum state of another ancillary qubit, represented by a orange blob. The influence of the CNOT and Hadamard gates projects the outcome of this stabilizer operation onto the ancillary qubit, altering its state. Subsequently, the modified state of the ancillary qubit is measured to ascertain the syndrome value S\textsubscript{x}. This setup not only allows for precise error detection but also ensures that the corrections needed can be accurately determined and applied to maintain the quantum system's overall fidelity.

In all cases, the interaction between the stabilizers and the ancillary qubits is critical for detecting errors effectively without altering the logical state of the quantum system. These visual and operational details highlight the intricate and precise nature of error correction strategies employed in quantum computing, specifically within the framework of surface codes. Although surface codes offer substantial benefits in error correction, their implementation demands a considerable number of physical qubits and sophisticated control systems. These requirements pose substantial hurdles in the development of a fault-tolerant quantum computer.

\subsection{Transversal Gates}

Surface codes are designed around a two-dimensional nearest-neighbor (2DNN) architecture, where qubits are arranged to interact primarily with their direct neighbors. This configuration aligns well with the physical layout of qubits on quantum chips, simplifying quantum operations and minimizing error probabilities due to its unity with the quantum chip's architecture. The inherent design of surface codes ensures that error correction is efficiently achieved through interactions among nearest-neighbor qubits. Despite this, there has been a proposal to execute multi-qubit gate operations transversally across different surface codes \cite{abobeih2022fault, freedman2001projective, bravyi1998quantum}. Simplifying the concept, if an initial circuit contains an $X$ gate, implementing it transversally means applying $X$ gates to every qubit in the lattice. Likewise, suppose the original circuit involves a $CNOT$ gate between two qubits. In that case, the transversal approach necessitates executing several $CNOT$ gates from all qubits in one encoded lattice to those in another. Figure \ref{fig:surface_code} \raisebox{.5pt}{\textcircled{\raisebox{-.9pt} {\textbf{5}}}} and \raisebox{.5pt}{\textcircled{\raisebox{-.9pt} {\textbf{6}}}} illustrates the application of 2-qubit interactions transversally from one lattice to another in both unrotated and rotated surface codes. 

Nonetheless, the necessity for transversal two-qubit gates has historically rendered planar encoding impractical in several scenarios where physical qubits are limited to 2D configurations and can only engage in nearest-neighbor interactions. This limitation is particularly evident in systems such as quantum dots \cite{jones2010layered, herrera2010photonic}, superconducting qubits \cite{divincenzo2009fault, groszkowski2011tunable}, trapped atoms \cite{kruse2010reconfigurable}, nitrogen-vacancy centers in diamonds \cite{yao2012scalable}, and certain ion trap setups \cite{kumph2, crick2010fast}. Implementing transversal gates in a surface code can challenge the 2DNN structure for several reasons: \circled {1} Breaking Locality: To perform a transversal gate, one needs to apply operations across potentially distant qubits simultaneously. In a strict 2D lattice, this means reaching beyond immediate neighbors, which disrupts the locality principle inherent to surface codes. Maintaining only nearest-neighbor interactions is crucial for minimizing error rates and implementation complexity. \circled{2} Increased Error Propagation Risk: While transversal gates are designed to prevent error propagation within their definition, the act of physically implementing these gates in a surface code setting—where we might have to engage non-neighbor qubits—increases the risk of spreading errors. This contradicts the surface code's principle of localizing errors for easier detection and correction. \circled{3} Implementation Complexity: The surface code's error correction relies on inherently local measurements. Introducing transversal gates necessitates control and synchronization across a wider array of qubits, complicating the implementation and potentially introducing more points of failure, which can degrade the error correction capabilities of the surface codes. \circled{4} Limited Set of Transversal Gates: Not all quantum gates can be implemented transversally in a way that maintains the 2DNN structure of surface codes. This limitation means that some desired quantum operations cannot be performed without compromising the local interaction model, thus forcing a trade-off between the types of operations one can perform and the preservation of the 2DNN structure.

In conclusion, although transversal gates contribute to fault tolerance, their integration within surface codes disrupts the 2DNN architecture by requiring interactions beyond immediate neighbors. This necessity introduces complexities in physical implementation, elevates the potential for error spreading and restricts the variety of operations that can be executed while adhering to the principle of locality. These issues have been addressed through the innovative approach of lattice surgery, which involves the strategic `cutting' and `merging' of code surfaces. This method effectively preserves the integrity of standard nearest-neighbor interactions and fault tolerance, offering a solution to the challenges posed by transversal gates within the framework of surface codes.

%% file: 3_lattice_surgery.tex
\section{Lattice Surgery} \label{lattice_surgery}

\subsection{Pivotal Idea}

Lattice Surgery involves the strategic manipulation of a lattice's structure to obtain specific outcomes, utilizing two fundamental techniques: merging and splitting. Merging two lattices entails combining them into one unified surface. This process is facilitated by the measurement of joint stabilizers along the surfaces' edges as part of error correction cycles. The behavior of the operation varies based on the edges that are connected. Conversely, splitting a lattice separates one surface into two by severing joint stabilizers, thereby creating additional boundaries. The characteristics of the newly formed boundaries dictate the properties of the resultant states. Every surface code has a rough edge and a smooth edge. They describe the boundaries of a two-dimensional lattice that encodes qubits. Visually, the rough edges are identified by the explicit termination of qubits at the lattice boundary. While, the smooth edges are defined by the absence of qubit termination, implying a conceptual continuation of the lattice structure beyond its physical confines. Rough edges, through the termination of qubits, provide a structural basis for the application of X-stabilizers. Smooth edges, characterized by the non-termination of qubits, facilitate the deployment of Z-stabilizers. The configuration of these qubits about the lattice's boundary defines the operational dynamics of the code, particularly in the correction of bit-flip and phase-flip errors.

\subsection{Lattice Merging}

\begin{figure}
    \centering
    \includegraphics[width=0.8\linewidth]{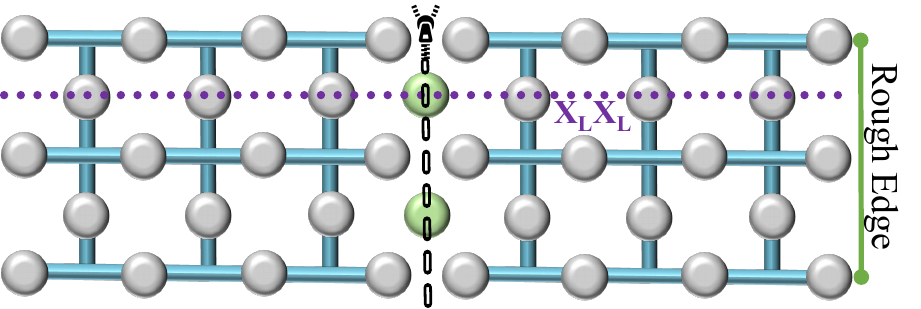}
    \caption{\textbf{Rough surface merging.} 
    Two logical surface codes are unified by a row of transitional qubits (green blobs) initialized in the $\ket{0}$ state, facilitating a merged surface encoded by the operation $X_LX_L$.
    }
    \label{fig:rough_lattice_merging}
\end{figure}

\begin{figure}
    \centering
    \includegraphics[width=0.8\linewidth]{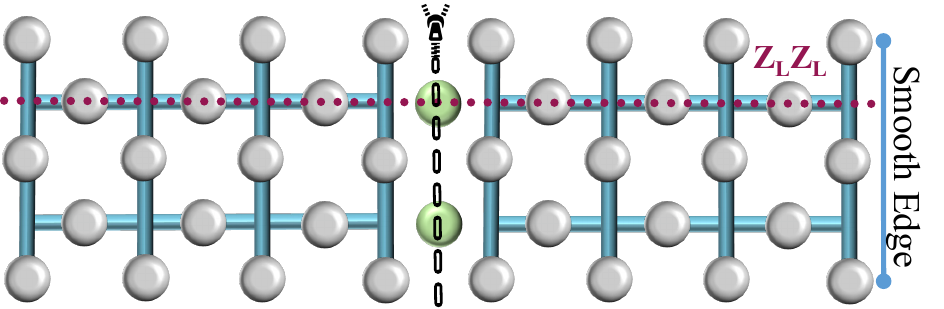}
    \caption{\textbf{Smooth surface merging.} 
    The unification of two logical surface codes through transitional qubits (yellow blobs) prepared in the $\ket{+}$ state, culminating in a merged surface characterized by $Z_LZ_L$.
    }
    \label{fig:smooth_lattice_merging}
\end{figure}

Consider the setup depicted in Fig. \ref{fig:rough_lattice_merging}, comprising two logical surface codes, each encoding a single qubit, and a separate row of `transitional' uninitialized physical qubits, marked in green. Initially, these qubits are set to the $\ket{0}$ state. Subsequently, the two surfaces are unified by positioning the transitional qubits between them. Additional stabilizers are created around the new qubits to incorporate them within the system. The entire system then undergoes error correction as a unified data surface. The logical operation X, denoted as the old boundary, remains unchanged and is represented by the product of the logical operators from the two initial surfaces, thus denoted as $X_LX_L$. Consequently, this merging process, which combines the rough edges of the two surfaces, is identified as rough surface merging, resulting in a single, unified surface. In the second merging technique, referred to as smooth surface merging (Fig. \ref{fig:smooth_lattice_merging}), the transitional qubits (depicted as green blobs) are initialized in the $\ket{+}$ state. The resulting merge is characterized by the measurement of $Z_LZ_L$, distinguishing it from the rough surface merging by the initial state of the transitional qubits and the nature of the logical operation measured post-merge.

\subsection{Lattice Splitting}

\begin{figure}
    \centering
    \includegraphics[width=0.8\linewidth]{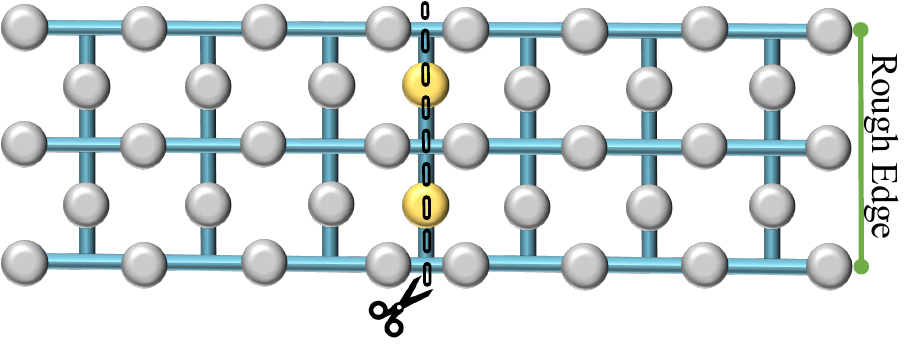}
    \caption{\textbf{Rough surface splitting.} 
    Division of a single quantum surface into two separate entities via the measurement of transitional qubits (yellow blobs).
    }
    \label{fig:rough_lattice_splitting}
\end{figure}

\begin{figure}
    \centering
    \includegraphics[width=0.8\linewidth]{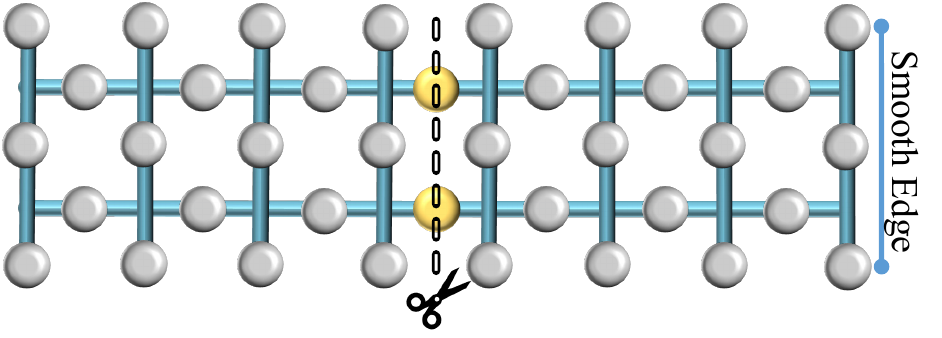}
    \caption{\textbf{Rough surface splitting.} 
    The central row of qubits is measured to create separate surfaces with distinct boundary conditions.
    }
    \label{fig:smooth_lattice_splitting}
\end{figure}

Lattice splitting, entails dividing a single surface into two by measuring a central row of qubits, effectively excising them from the lattice. This measurement results in the formation of two distinct surfaces upon completion of the operation. Similar to merging, splitting can occur along two types of boundaries: rough or smooth. In the case of a rough split, illustrated in Fig. \ref{fig:rough_lattice_splitting}, the central row of qubits—visually denoted as yellow blobs — are measured in the Pauli-X basis, effectively partitioning the surface into two independent lattices. This division can alter the code's distance, potentially affecting its error-correcting capabilities. Likewise, during a smooth split, as depicted in Fig. \ref{fig:smooth_lattice_splitting}, the central row of intermediate qubits—again represented as yellow blobs—is similarly measured out with similar effect as rough split. 

%% file: 4_gate_operations.tex
\section{Operations with Lattice Surgery} \label{gate_operations}

In this section, we describe the emulation of fundamental gate operations using lattice surgery. Familiarity with the logical $X$ and $Z$ behaviors is extended to examine additional gates. 

\subsection{The CNOT Gate}

\begin{figure}
    \centering
    \includegraphics[width=1\linewidth]{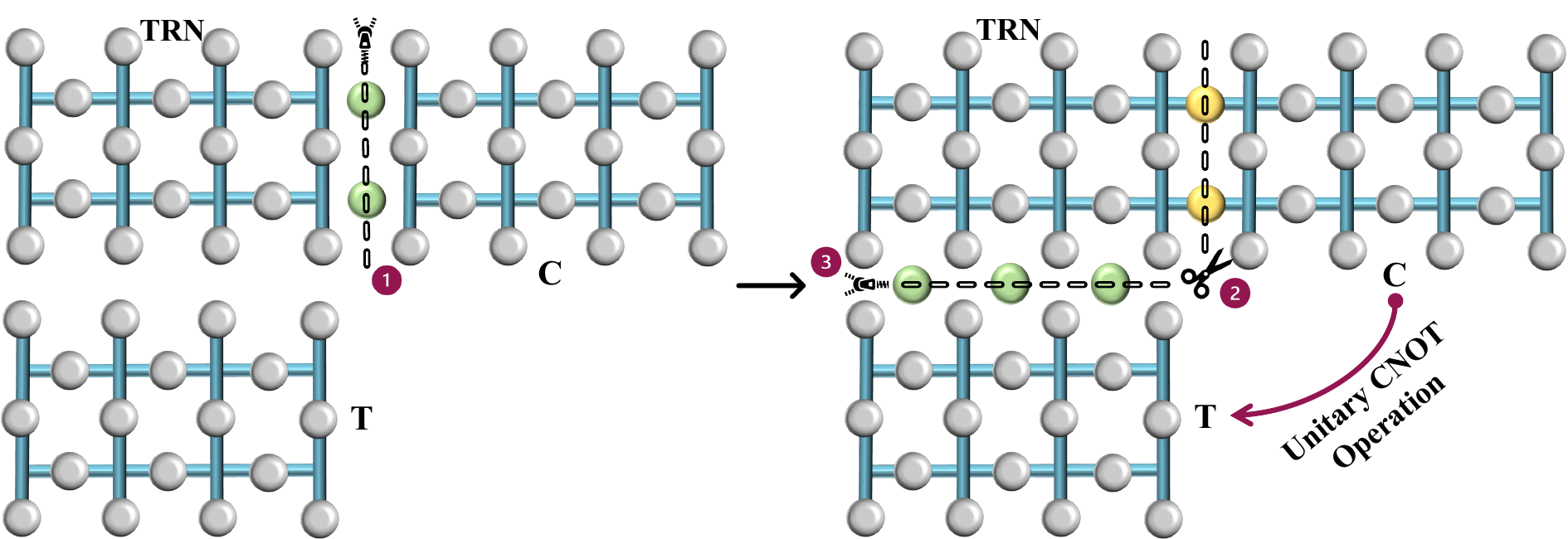}
    \caption{\textbf{Implementation of the $CNOT$ gate using lattice surgery.} 
    The process involves merging and splitting logical qubits $C$ (control) and $T$ (target) with a transitional surface $TRN$, strategically applying quantum operations to achieve the gate function within a surface code environment.
    }
    \label{fig:cnot}
\end{figure}

The $CNOT$ gate creation process is a technique used to perform logical operations between qubits encoded in surface codes. The entire process is shown in Fig. \ref{fig:cnot}. We begin with with two logical qubits of distance $d$, the control ($C$) and the target ($T$), each encoded on separate surfaces. The control is in an arbitrary state while the target is initialized in the state $\ket{+}$. Just as ancilla qubits are utilized, we introduce an auxiliary surface known as the transitional logical surface ($TRN$). The $TRN$ mirrors the structure of the surfaces of $C$ and $T$, and is therefore a surface with a distance $d$. This entire surface is initialized in the $\ket{+}$ state at the start of the operation. This $TRN$ surface is used to bridge the control and target qubits and facilitate the $CNOT$ operation.

The first operation is a smooth merge between the surfaces $C$ and $TRN$. The smooth merge is used here because it conserves the phase relationship, which is necessary for the control functionality of the $CNOT$ gate. After the merge, a logical state dependent on the measurement outcome is formed. The merged surface is then split back into $C$ and $TRN$. This split is also a smooth split. The purpose of the split is to `separate' the control and the intermediary while preserving the state information transferred during the merge. Finally, the transitional $TRN$ is merged with the target $T$. This merge effectively performs the controlled-NOT operation: if the control qubit was in the state $\ket{1}$, the target's state will be flipped. If the control was in $\ket{0}$, the target remains unchanged. After the final merge between $TRN$ and $T$, we achieve the $CNOT$ operation between the control and target qubits. In the end, we are left with two logical qubits of distance $d$. The $CNOT$ operation is also completely reversible.

The series of merges and splits are not redundant but rather a systematic way to transfer and manipulate the quantum information between the qubits to realize the $CNOT$ gate. Each merge and split has a purpose: The first merge is to entangle the control qubit's state with the intermediary. The split is necessary to retain the control's state while preparing to apply its effect to the target. The second merge then applies the control qubit's state to the target qubit, completing the $CNOT$ operation. The process exploits the properties of quantum mechanics to implement computational logic in a fundamentally different way than classical logic gates. After completing this series of operations, the $TRN$ becomes inactive, freeing the qubits for further use. We can either reinitialize the surface to the $\ket{+}$ state for a forthcoming $CNOT$ operation in the circuit, or reset the qubits for a different purpose.

\subsection{The Hadamard Gate}

\begin{figure}
    \centering
    \includegraphics[width=0.7\linewidth]{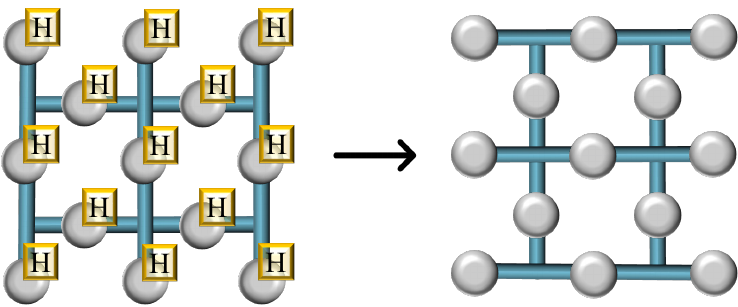}
    \caption{\textbf{Orientation disruption post-transversal Hadamard gate.} 
    Resulting configuration after applying a transversal Hadamard gate: The individual $H$ operations reorient the planar surface, diverging from its original layout as demonstrated in this figure.
    }
    \label{fig:H_rotate}
\end{figure}

Executing a Hadamard gate across an entire surface of qubits transversally, through the application of the $H$ operation to each qubit individually, results in a planar surface that is correct state-wise. However, this approach alters the physical orientation of the planar surface relative to its initial configuration, as depicted in Fig. \ref{fig:H_rotate}. Such reorientation is acceptable if the qubits require no additional manipulation. Nonetheless, should subsequent operations be necessary, or if the preservation of the qubits' interconnectivity is essential for scaling the system, an alternative strategy must be employed to realign the planar surface to its original layout.

\begin{figure}
    \centering
    \includegraphics[width=0.8\linewidth]{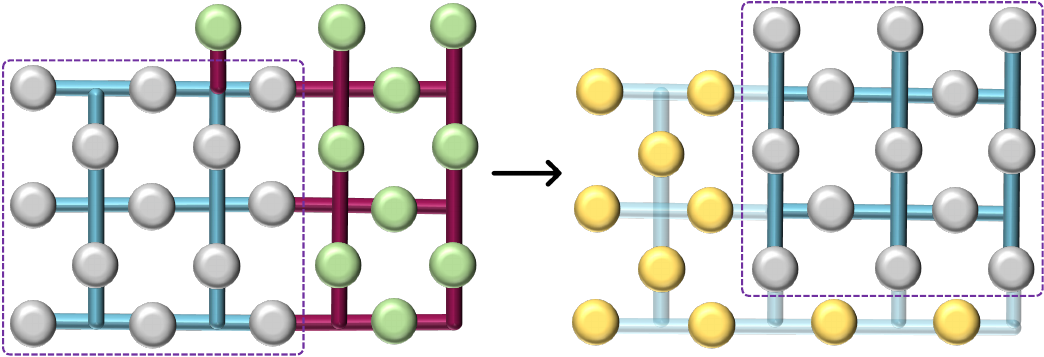}
    \caption{\textbf{Restoration of planar surface orientation.} 
    Corrective process to realign the planar surface following a transversal Hadamard operation: Merging with auxiliary qubits and subsequent contraction through Z-basis measurements returns the surface to its original orientation, as depicted here.
    }
    \label{fig:H_fix_rotate_back}
\end{figure}

To realign the planar surface post-Hadamard operation and preserve its original orientation—necessary for further quantum computations or to maintain the qubits' connectivity for scalable quantum systems—specific corrective steps can be undertaken, as outlined in Fig. \ref{fig:H_fix_rotate_back}. The process begins with the rotated surface, which is subsequently merged with additional qubits (depicted as green blobs). This merger creates an enlarged, stable surface, which compensates for the reorientation caused by the initial Hadamard operations. Subsequently, this expanded surface is methodically reduced back to its original size by measuring certain qubits (marked in yellow) in the Z-basis, effectively reversing the rotation. Through this series of splits, the planar surface's alignment is restored. This corrective sequence of merges and splits functions due to the topological nature of surface codes, where logical operations like the Hadamard can be mimicked by altering the connectivity and layout of the qubits. The final outcome, after contracting the surface, is a qubit array that has effectively undergone a Hadamard transformation and is realigned to its initial configuration, ready for subsequent quantum processing steps.

\subsection{Arbitrary Qubit Rotation Gates}

\begin{figure}
    \centering
    \includegraphics[width=0.8\linewidth]{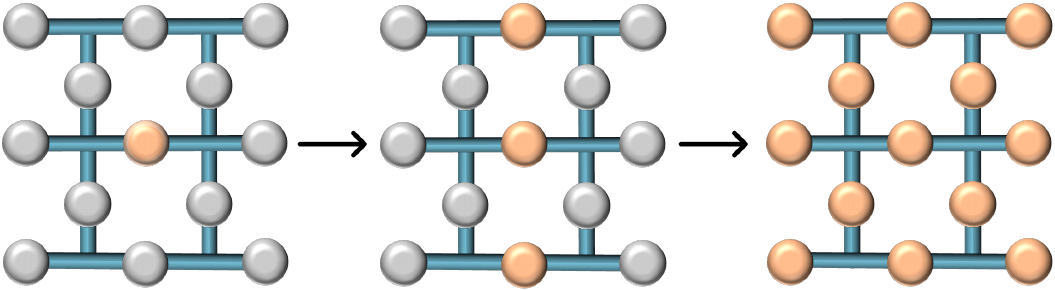}
    \caption{\textbf{State injection into a quantum lattice.} 
    Sequential steps illustrating the injection of an arbitrary quantum state $\Phi$ into a lattice: starting with initialization, performing $CNOT$ operations, swapping syndrome with data qubits, and spreading the state to achieve $\alpha' \ket{0}_L + \beta' \ket{1}_L$.
    }
    \label{fig:arbit_qubit_rotation}
\end{figure}

\begin{figure}
    \centering
    \includegraphics[width=0.8\linewidth]{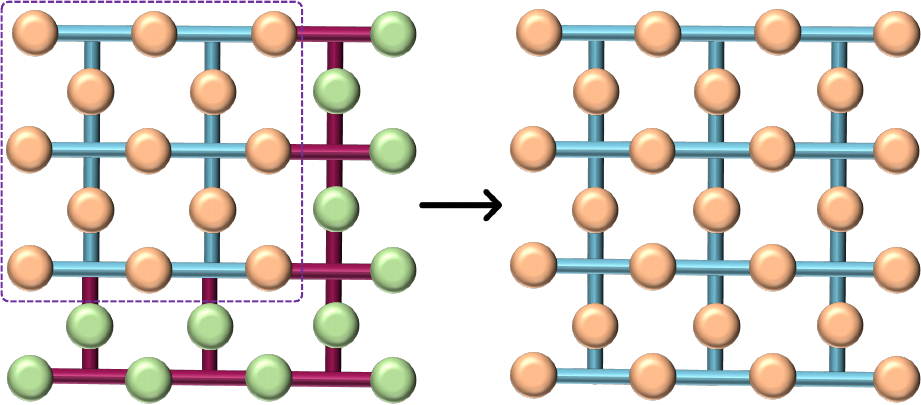}
    \caption{\textbf{Scaling the injected state in a higher distance lattice.} 
    The process of expanding an injected state across a larger lattice to form a high-distance logical state, depicted through the inclusion of additional $\ket{0}$ qubits and stabilization to reach $\alpha' \ket{0}_{L'} + \beta' \ket{1}_{L'}$.
    }
    \label{fig:arbit_qubit_rotation_scaling}
\end{figure}

Having grasped the principles of integrating $CNOT$ and Hadamard gates within lattice structures, we can now extend our exploration to the technique of state injection across a whole lattice. Consider a quantum state $\Psi = \alpha \ket{0} + \beta \ket{1}$. The application of an arbitrary quantum gate to this state effectively `rotates' it, transitioning it to a new state $\Phi = \alpha' \ket{0} + \beta' \ket{1}$. This transformation is due to the fact that all quantum gates act as rotations in the state space. While transversal operations could be employed to achieve this effect, they would only result in an altered orientation of the lattice's surface. To incorporate a specific arbitrary state into a given lattice, and to accomplish the desired quantum gate's effect without altering the lattice orientation, we employ lattice surgery. This technique allows for the precise and controlled introduction of the rotated state into the quantum computing framework.

The procedure for state injection is illustrated in Fig. \ref{fig:arbit_qubit_rotation}. To inject a state $\Phi$ into a lattice, we begin with all qubits in the lattice initialized to the state $\ket{0}$, except for one, which is in the state $\Phi$. As shown in the first panel of the diagram, the grey qubits represent those in the state $\ket{0}$, and the orange qubit is in the state $\Phi$. CNOT operations are conducted between the orange qubit and the adjacent syndrome qubits. These syndrome qubits are subsequently swapped with the neighboring data qubits to create a three-qubit state $\alpha' \ket{000} + \beta' \ket{111}$, as indicated in the second panel of Fig. \ref{fig:arbit_qubit_rotation}. The lattice undergoes routine stabilizer operation, effectively spreading the encoded state across the entire structure, as depicted in the third step of Fig. \ref{fig:arbit_qubit_rotation}. This distribution is facilitated by syndrome measurements, which function as repeated applications of $CNOT$ gates. This series of operations culminates in the injection of the state onto the entire lattice, through progressive entanglement, yielding a new logical state $\alpha' \ket{0}_L + \beta' \ket{1}_L$.

\begin{figure*}
    \centering
    \includegraphics[width=1\linewidth]{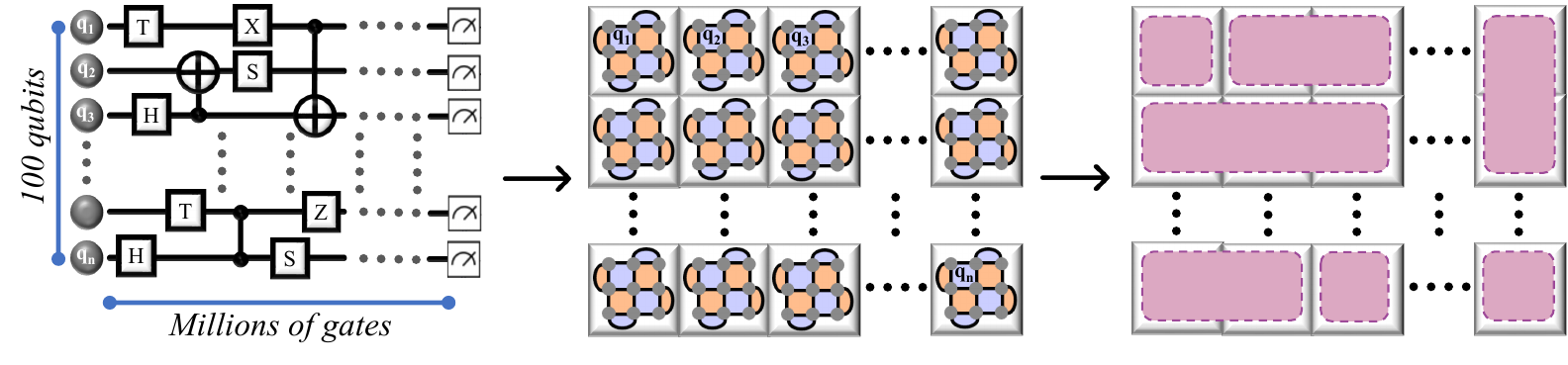}
    \caption{\textbf{Integration of QECC in large-scale quantum circuits.} 
    Depiction of a complex quantum circuit comprising $100$ qubits, each represented by individual surface codes in a tiled layout. This diagram illustrates how each tile (or surface) interacts with others through operations like merging, splitting, expanding, and contracting to faithfully emulate the gate operations of the original quantum circuit, ensuring fault tolerance in large-scale applications.
    }
    \label{fig:100_qubits_qecc}
\end{figure*}

Let us imagine a single qubit influenced by one rotational gate, $R_x(\theta)$. This qubit is allocated its own dedicated surface. We then select a single qubit from this surface and apply the same rotational gate, $R_x(\theta)$, as used on the original qubit. Following the procedure outlined in \ref{fig:arbit_qubit_rotation}, we proceed with the stabilizer operations or CNOT gates until the entire surface replicates the effect of the rotational X gate on the qubit. Once an injected surface is established, it can be scaled up to a surface with a higher distance. This scalable approach extends the injected state $\Phi$ across a larger lattice to achieve a higher error-correcting distance. As illustrated in Fig. \ref{fig:arbit_qubit_rotation_scaling}, the expansion of the injected state onto a broader lattice is facilitated through the incorporation of extra qubits (marked in green) initialized in the state $\ket{0}$. Subsequent rounds of stabilizer measurements are then performed, allowing the state $\Phi$ to permeate the enlarged lattice. This process culminates in the formation of an augmented logical state, represented as $\alpha' \ket{0}_{L'} + \beta' \ket{1}_{L'}$, across the new, higher distance surface.

%% file: 5_multi_qubit.tex
\section{Application of QECC on multi-qubit circuits} \label{multi_qubit}

\begin{figure*}
    \centering
    \includegraphics[width=1\linewidth]{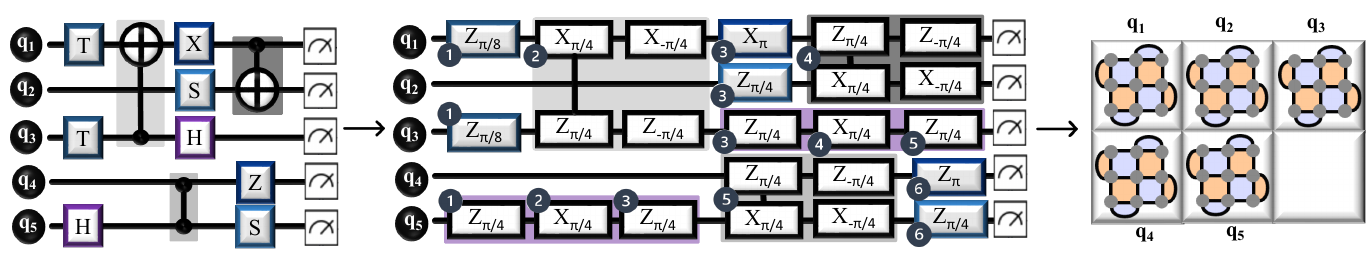}
    \caption{\textbf{Quantum circuit decomposition and surface code mapping.} 
    Illustration of the process of decomposing a complex quantum circuit (first circuit) into its Pauli rotational forms (second circuit) and the subsequent integration with lattice surgery through surface codes (third diagram). Each qubit in the circuit is associated with a surface code, shown here as distance 3 rotated codes.
    }
    \label{fig:simplify_circuit}
\end{figure*}

\begin{figure*}
    \centering
    \includegraphics[width=1\linewidth]{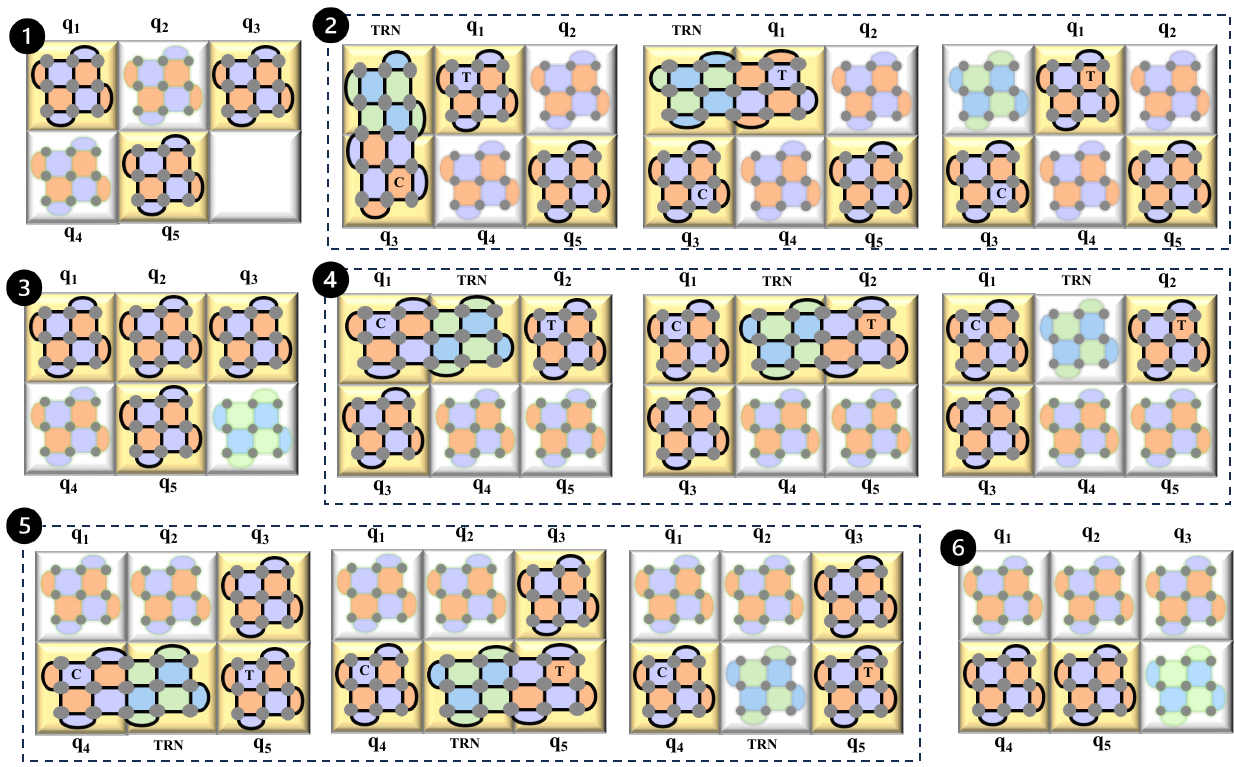}
    \caption{\textbf{Detailed steps of lattice surgery in a quantum circuit.} 
    This figure demonstrates the step-by-step integration of lattice surgery techniques in a decomposed quantum circuit (second circuit of Fig. \ref{fig:simplify_circuit}). Starting with state injections for the $T$ and $H$ gates in the first step, the diagram progresses through the creation of $CNOT$ and $CZ$ gates, and concludes with logical operations on specific qubits. Each step corresponds to the placement of state injections and logical gate operations, illustrating how lattice surgery manipulates and measures stabilizers to maintain the fidelity of the quantum state throughout the process. 
    }
    \label{fig:all_steps}
\end{figure*}

\subsection{Moving Towards Fault-Tolerant Quantum Computing}

We have explored how lattice surgery facilitates quantum operations, focusing primarily on single-qubit tasks and extending up to two-qubit computations, such as implementing a $CNOT$ gate between two-qubit lattices. However, as quantum computing scales up to include circuits with a larger number of qubits, the complexity of applications also grows, demanding numerous qubits and gates. To achieve fault tolerance in these expansive applications, utilizing multiple QECC surfaces, each corresponding to individual qubits within the circuit is necessary. 

This multi-surface approach is just the beginning. There is also a critical need to orchestrate interactions among these QECC surfaces to accurately replicate the gate operations found in the original quantum circuit. Figure \ref{fig:100_qubits_qecc} illustrates an arbitrary quantum circuit with $100$ qubits, encompassing millions of gates. Here, each qubit is represented by a distinct tile-like surface code, where each tile corresponds to a logical surface associated with specific qubits. The interaction between these tiles — merging, splitting, expanding, or contracting — is crucial for emulating the intended gate operations within the circuit, ensuring that the larger-scale quantum computing applications remain robust and error-resistant. Executing a quantum circuit on a surface-code-based quantum computer as efficiently as possible involves an optimization challenge. This challenge, which has been proven to be NP-hard \cite{herr2017optimization}, focuses on minimizing both the number of surface code tiles and the number of time steps required to implement quantum algorithms. The objective is to achieve the most efficient use of resources in terms of both space and time.

\subsection{Simplification of the Circuit}

\input{q_gate_table}

One effective method to sequentially execute all interactions from the original circuit is to decompose them into simpler subproblems. We recognize that each quantum gate can be expressed in terms of a Pauli rotation, denoted as $P_\phi = e^{-iP\phi}$, where $P$ represents a Pauli product operator such as $Z$, $X$, $Y$ or $Y \otimes X$, among others and $\phi$ is an angle \cite{litinski2019game}. Table \ref{tab:gates_pauli_rotation_table} provides a list of quantum gates along with their corresponding forms in Pauli rotations. By utilizing these rotational forms, we can simplify a quantum circuit, which facilitates the application of techniques like lattice surgery later on. To illustrate how lattice surgery can be effectively integrated into a quantum circuit, let us examine a specific example, as depicted in Fig. \ref{fig:simplify_circuit}. This example circuit comprises various single-qubit and multi-qubit gates. The initial step involves decomposing all the circuit's gates into their Pauli rotational forms. This decomposition is crucial as it facilitates the injection of quantum states into the appropriate surface codes where needed. The decomposed version of this circuit is presented in the second figure of Fig. \ref{fig:simplify_circuit}. Since the circuit involves five qubits, there will correspondingly be five surface codes, each linked to one of the qubits. The surface codes illustrated here are distance 3 rotated codes, although the specific type of code used can vary based on the physical error characteristics of the quantum system.

After decomposing the circuit, we describe lattice surgery which is a six step process as showcased in Fig. \ref{fig:all_steps}. At each stage of the process, the yellow tiles indicate the active surfaces involved in the operation of that specific step, while the other blurred surfaces represent those that are inactive during that step. The first step starts with a state injection into the qubit, $q_1$ and $q_2$ for the $T$ gates and another state injection into qubit, $q_5$ for the first decomposed gate of $H$ gate. The second step creates the $CNOT$ gate between the lattices of qubits, $q_1$ and $q_3$. For this we need to initialize a transitional lattice marked as $TRN$. To emulate the $CNOT$ gate, the $TRN$ is initially merged with qubit $q_3$. Subsequently, they are separated, allowing the $TRN$ to then merge with qubit $q_1$. In the final arrangement, $q_3$ serves as the control and $q_1$ as the target, while the $TRN$ becomes inactive. Although the $TRN$ qubits can be repurposed for other uses, in this example we will reinitialize the $TRN$ surface to the $\ket{+}$ state each time it becomes inactive. This ensures that the surface is ready to function as a $TRN$ surface for subsequent multi-qubit operations. The third step contains four substeps - the first applies a logical $X$ gate on the lattice of qubit, $q_1$, then it proceeds to three state injections - into the lattice of $q_2$ for the $S$ gate, then into $q_3$ for the first decomposed gate of the $H$ gate and lastly on $q_5$ for the last decomposed gate of the $H$ gate. The fourth step mimics the $CNOT$ gate between the lattices of qubits, $q_1$ and $q_2$, by using a transitional lattice marked as $TRN$, followed by a state injection into $q_3$ for the second decomposed gate of the $H$ gate. To simulate the $CNOT$ gate, the $TRN$ first merges with qubit $q_1$. After this, the $TRN$ and $q_1$ are separated, which then allows the $TRN$ to merge with qubit $q_2$. In the concluding setup, $q_1$ acts as the control qubit and $q_2$ as the target, while the $TRN$ becomes inactive. The fifth step establishes a $CZ$ gate between the lattices of $q_4$ and $q_5$ making use of the transitional lattice $TRN$. To replicate the CZ gate, the $TRN$ initially combines with qubit $q_4$. They are then separated, enabling the $TRN$ to subsequently merge with qubit $q_5$. In this final configuration $TRN$ is rendered inactive. This step also involves a state injection on qubit, $q_3$ for the last decomposed gate of the $H$ gate. The final or the sixth step involves applying a logical $Z$ operator onto qubit, $q_4$ and a final state injection onto qubit, $q_5$ for the last $S$ gate. With this we finish all the steps of lattice surgery and the final stabilizers after all sorts of merging, splitting, expanding and contracting, that can be measured, mimics the true nature of the original five qubit circuit.

Lattice surgery, as depicted for smaller quantum circuits, scales up to encompass systems with hundreds of qubits, marking a critical progression towards fault-tolerant quantum computing. In large-scale implementations, lattice surgery involves the intricate manipulation of a vast network of qubits, each encoded and interconnected through surface codes or other error-correcting codes to form a robust quantum lattice. By extending the principles observed in smaller setups—such as sequential state injections, logical operations mapped across multiple qubits, and adaptive merging and splitting of stabilizers—lattice surgery on hundreds of qubits allows for complex computational tasks that are beyond the reach of classical computers. The modular nature of lattice surgery aids in efficiently managing quantum resources, enabling selective entanglement and disentanglement of qubits as required by the algorithmic demands. Furthermore, the ability to execute error correction through merging and splitting surface code patches without the need to physically relocate qubits or disrupt the entire quantum state highlights the practicality of lattice surgery in large-scale applications. 

%% file: q_gate_table.tex
\begin{table}[]
\centering
\caption{Quantum Gates and their Pauli Rotational Form}
\begin{tabular}{cc||c}
\multicolumn{2}{c||}{\textbf{Quantum Gates}}                                                                                          & \textbf{Rotational Form} \\ \hline \hline
\multicolumn{1}{c|}{\multirow{9}{*}{\textbf{\begin{tabular}[c]{@{}c@{}}Single\\ Qubit\end{tabular}}}} & \textbf{X}           & $X_{\pi}$                       \\
\multicolumn{1}{c|}{}                                                                                         & \textbf{Y}           & $Y_{\pi}$                       \\
\multicolumn{1}{c|}{}                                                                                         & \textbf{Z}           & $Y_{\pi}$                       \\
\multicolumn{1}{c|}{}                                                                                         & \textbf{RX}          & $X_{\theta}$                      \\
\multicolumn{1}{c|}{}                                                                                         & \textbf{RY}          & $Y_{\theta}$                      \\
\multicolumn{1}{c|}{}                                                                                         & \textbf{RZ}          & $Z_{\theta}$                      \\
\multicolumn{1}{c|}{}                                                                                         & \textbf{H}           & $Z_{\pi /4} \cdot X_{\pi /4} \cdot Z_{\pi /4}$                       \\
\multicolumn{1}{c|}{}                                                                                         & \textbf{S}           & $Z_{\pi /4}$                       \\
\multicolumn{1}{c|}{}                                                                                         & \textbf{T}           & $Z_{\pi /8}$                       \\ \hline \hline
\multicolumn{1}{c|}{\multirow{3}{*}{\textbf{\begin{tabular}[c]{@{}c@{}} \\Multi\\ Qubit\end{tabular}}}}  & & \\

\multicolumn{1}{c|}{} & \textbf{CNOT}        & $(Z \otimes X)_{\pi /4} \cdot (\mathds{1} \otimes X)_{-\pi /4} \cdot (Z \otimes \mathds{1})_{-\pi /4}$                    \\
\multicolumn{1}{c|}{}                                                                                         & \textbf{C(P1, P2)}   & $(P_1 \otimes P_2)_{\pi /4} \cdot (\mathds{1} \otimes P_1)_{-\pi /4} \cdot (P_2 \otimes \mathds{1})_{-\pi /4}$                       \\
\multicolumn{1}{c|}{}                                                                                         &  &                      \\ \hline \hline
\end{tabular}
\label{tab:gates_pauli_rotation_table}
\end{table}

%% file: 7_conclusion.tex
\section{Conclusion} \label{conclusion}

This paper attempts to simplify the complex topic of lattice surgery in the field of fault tolerant quantum computing.
It explores surface codes and demonstrates the practical application of lattice surgery in constructing quantum gates and simulate multi-qubit circuits to address this complex subject. 
As quantum computing scales to encompass larger and more complex systems, lattice surgery provides a critical framework 
, thereby enhancing the robustness and operational capabilities of quantum computers. The continued evolution of this field will likely see lattice surgery not only adapting to new quantum architectures but also inspiring analogous methodologies across various error-correcting paradigms, which will be crucial for achieving scalable and fault-tolerant quantum computing.

%% file: 0_main.bbl
\begin{thebibliography}{10}
\providecommand{\url}[1]{#1}
\csname url@samestyle\endcsname
\providecommand{\newblock}{\relax}
\providecommand{\bibinfo}[2]{#2}
\providecommand{\BIBentrySTDinterwordspacing}{\spaceskip=0pt\relax}
\providecommand{\BIBentryALTinterwordstretchfactor}{4}
\providecommand{\BIBentryALTinterwordspacing}{\spaceskip=\fontdimen2\font plus
\BIBentryALTinterwordstretchfactor\fontdimen3\font minus \fontdimen4\font\relax}
\providecommand{\BIBforeignlanguage}[2]{{%
\expandafter\ifx\csname l@#1\endcsname\relax
\typeout{** WARNING: IEEEtran.bst: No hyphenation pattern has been}%
\typeout{** loaded for the language `#1'. Using the pattern for}%
\typeout{** the default language instead.}%
\else
\language=\csname l@#1\endcsname
\fi
#2}}
\providecommand{\BIBdecl}{\relax}
\BIBdecl

\bibitem{reiher2017elucidating}
M.~Reiher, N.~Wiebe, K.~M. Svore, D.~Wecker, and M.~Troyer, ``Elucidating reaction mechanisms on quantum computers,'' \emph{Proceedings of the national academy of sciences}, vol. 114, no.~29, pp. 7555--7560, 2017.

\bibitem{orus2019quantum}
R.~Or{\'u}s, S.~Mugel, and E.~Lizaso, ``Quantum computing for finance: Overview and prospects,'' \emph{Reviews in Physics}, vol.~4, p. 100028, 2019.

\bibitem{schuld2015introduction}
M.~Schuld, I.~Sinayskiy, and F.~Petruccione, ``An introduction to quantum machine learning,'' \emph{Contemporary Physics}, vol.~56, no.~2, pp. 172--185, 2015.

\bibitem{gachnang2022quantum}
P.~Gachnang, J.~Ehrenthal, T.~Hanne, and R.~Dornberger, ``Quantum computing in supply chain management state of the art and research directions,'' \emph{Asian Journal of Logistics Management}, vol.~1, no.~1, pp. 57--73, 2022.

\bibitem{ajagekar2019quantum}
A.~Ajagekar and F.~You, ``Quantum computing for energy systems optimization: Challenges and opportunities,'' \emph{Energy}, vol. 179, pp. 76--89, 2019.

\bibitem{clerk2010introduction}
A.~A. Clerk, M.~H. Devoret, S.~M. Girvin, F.~Marquardt, and R.~J. Schoelkopf, ``Introduction to quantum noise, measurement, and amplification,'' \emph{Reviews of Modern Physics}, vol.~82, no.~2, p. 1155, 2010.

\bibitem{mouloudakis2021entanglement}
G.~Mouloudakis and P.~Lambropoulos, ``Entanglement instability in the interaction of two qubits with a common non-markovian environment,'' \emph{Quantum Information Processing}, vol.~20, pp. 1--15, 2021.

\bibitem{lo1998introduction}
H.-K. Lo, T.~Spiller, and S.~Popescu, \emph{Introduction to quantum computation and information}.\hskip 1em plus 0.5em minus 0.4em\relax World Scientific, 1998.

\bibitem{devitt2013quantum}
S.~J. Devitt, W.~J. Munro, and K.~Nemoto, ``Quantum error correction for beginners,'' \emph{Reports on Progress in Physics}, vol.~76, no.~7, p. 076001, 2013.

\bibitem{hamming1950error}
R.~W. Hamming, ``Error detecting and error correcting codes,'' \emph{The Bell system technical journal}, vol.~29, no.~2, pp. 147--160, 1950.

\bibitem{wootters2009no}
W.~K. Wootters and W.~H. Zurek, ``The no-cloning theorem,'' \emph{Physics Today}, vol.~62, no.~2, pp. 76--77, 2009.

\bibitem{von2018mathematical}
J.~Von~Neumann, \emph{Mathematical foundations of quantum mechanics: New edition}.\hskip 1em plus 0.5em minus 0.4em\relax Princeton university press, 2018, vol.~53.

\bibitem{sundaresan2022matching}
N.~Sundaresan, T.~J. Yoder, Y.~Kim, M.~Li, E.~H. Chen, G.~Harper, T.~Thorbeck, A.~W. Cross, A.~D. C{\'o}rcoles, and M.~Takita, ``Matching and maximum likelihood decoding of a multi-round subsystem quantum error correction experiment,'' \emph{arXiv preprint arXiv:2203.07205}, 2022.

\bibitem{abobeih2022fault}
M.~Abobeih, Y.~Wang, J.~Randall, S.~Loenen, C.~Bradley, M.~Markham, D.~Twitchen, B.~Terhal, and T.~Taminiau, ``Fault-tolerant operation of a logical qubit in a diamond quantum processor,'' \emph{Nature}, vol. 606, no. 7916, pp. 884--889, 2022.

\bibitem{bacon2006operator}
D.~Bacon, ``Operator quantum error-correcting subsystems for self-correcting quantum memories,'' \emph{Physical Review A}, vol.~73, no.~1, p. 012340, 2006.

\bibitem{kitaev1997quantum}
A.~Y. Kitaev, ``Quantum computations: algorithms and error correction,'' \emph{Russian Mathematical Surveys}, vol.~52, no.~6, p. 1191, 1997.

\bibitem{krinner2022realizing}
S.~Krinner, N.~Lacroix, A.~Remm, A.~Di~Paolo, E.~Genois, C.~Leroux, C.~Hellings, S.~Lazar, F.~Swiadek, J.~Herrmann \emph{et~al.}, ``Realizing repeated quantum error correction in a distance-three surface code,'' \emph{Nature}, vol. 605, no. 7911, pp. 669--674, 2022.

\bibitem{bombin2006topological}
H.~Bombin and M.~A. Martin-Delgado, ``Topological quantum distillation,'' \emph{Physical review letters}, vol.~97, no.~18, p. 180501, 2006.

\bibitem{horsman2012surface}
D.~Horsman, A.~G. Fowler, S.~Devitt, and R.~Van~Meter, ``Surface code quantum computing by lattice surgery,'' \emph{New Journal of Physics}, vol.~14, no.~12, p. 123011, 2012.

\bibitem{landahl2014quantum}
A.~J. Landahl and C.~Ryan-Anderson, ``Quantum computing by color-code lattice surgery,'' \emph{arXiv preprint arXiv:1407.5103}, 2014.

\bibitem{cowtan2022qudit}
A.~Cowtan, ``Qudit lattice surgery,'' \emph{arXiv preprint arXiv:2204.13228}, 2022.

\bibitem{erhard2021entangling}
A.~Erhard, H.~Poulsen~Nautrup, M.~Meth, L.~Postler, R.~Stricker, M.~Stadler, V.~Negnevitsky, M.~Ringbauer, P.~Schindler, H.~J. Briegel \emph{et~al.}, ``Entangling logical qubits with lattice surgery,'' \emph{Nature}, vol. 589, no. 7841, pp. 220--224, 2021.

\bibitem{fowler2018low}
A.~G. Fowler and C.~Gidney, ``Low overhead quantum computation using lattice surgery,'' \emph{arXiv preprint arXiv:1808.06709}, 2018.

\bibitem{litinski2019game}
D.~Litinski, ``A game of surface codes: Large-scale quantum computing with lattice surgery,'' \emph{Quantum}, vol.~3, p. 128, 2019.

\bibitem{herr2018lattice}
D.~Herr, A.~Paler, S.~J. Devitt, and F.~Nori, ``Lattice surgery on the raussendorf lattice,'' \emph{Quantum Science and Technology}, vol.~3, no.~3, p. 035011, 2018.

\bibitem{chamberland2022circuit}
C.~Chamberland and E.~T. Campbell, ``Circuit-level protocol and analysis for twist-based lattice surgery,'' \emph{Physical Review Research}, vol.~4, no.~2, p. 023090, 2022.

\bibitem{chamberland2022universal}
------, ``Universal quantum computing with twist-free and temporally encoded lattice surgery,'' \emph{PRX Quantum}, vol.~3, no.~1, p. 010331, 2022.

\bibitem{herr2017optimization}
D.~Herr, F.~Nori, and S.~J. Devitt, ``Optimization of lattice surgery is np-hard,'' \emph{Npj quantum information}, vol.~3, no.~1, p.~35, 2017.

\bibitem{de2020zx}
N.~de~Beaudrap and D.~Horsman, ``The zx calculus is a language for surface code lattice surgery,'' \emph{Quantum}, vol.~4, p. 218, 2020.

\bibitem{vuillot2019code}
C.~Vuillot, L.~Lao, B.~Criger, C.~G. Almud{\'e}ver, K.~Bertels, and B.~M. Terhal, ``Code deformation and lattice surgery are gauge fixing,'' \emph{New Journal of Physics}, vol.~21, no.~3, p. 033028, 2019.

\bibitem{nielsen2001quantum}
M.~A. Nielsen and I.~L. Chuang, \emph{Quantum computation and quantum information}.\hskip 1em plus 0.5em minus 0.4em\relax Cambridge university press Cambridge, 2001, vol.~2.

\bibitem{chatterjee2023quantum}
A.~Chatterjee, K.~Phalak, and S.~Ghosh, ``Quantum error correction for dummies,'' in \emph{2023 IEEE International Conference on Quantum Computing and Engineering (QCE)}, vol.~1.\hskip 1em plus 0.5em minus 0.4em\relax IEEE, 2023, pp. 70--81.

\bibitem{chiaverini2004realization}
J.~Chiaverini, D.~Leibfried, T.~Schaetz, M.~D. Barrett, R.~Blakestad, J.~Britton, W.~M. Itano, J.~D. Jost, E.~Knill, C.~Langer \emph{et~al.}, ``Realization of quantum error correction,'' \emph{Nature}, vol. 432, no. 7017, pp. 602--605, 2004.

\bibitem{dennis2002topological}
E.~Dennis, A.~Kitaev, A.~Landahl, and J.~Preskill, ``Topological quantum memory,'' \emph{Journal of Mathematical Physics}, vol.~43, no.~9, pp. 4452--4505, 2002.

\bibitem{fowler2012surface}
A.~G. Fowler, M.~Mariantoni, J.~M. Martinis, and A.~N. Cleland, ``Surface codes: Towards practical large-scale quantum computation,'' \emph{Physical Review A}, vol.~86, no.~3, p. 032324, 2012.

\bibitem{kolmogorov2009blossom}
V.~Kolmogorov, ``Blossom v: a new implementation of a minimum cost perfect matching algorithm,'' \emph{Mathematical Programming Computation}, vol.~1, pp. 43--67, 2009.

\bibitem{tomita2014low}
Y.~Tomita and K.~M. Svore, ``Low-distance surface codes under realistic quantum noise,'' \emph{Physical Review A}, vol.~90, no.~6, p. 062320, 2014.

\bibitem{freedman2001projective}
M.~H. Freedman and D.~A. Meyer, ``Projective plane and planar quantum codes,'' \emph{Foundations of Computational Mathematics}, vol.~1, pp. 325--332, 2001.

\bibitem{bravyi1998quantum}
S.~B. Bravyi and A.~Y. Kitaev, ``Quantum codes on a lattice with boundary,'' \emph{arXiv preprint quant-ph/9811052}, 1998.

\bibitem{jones2010layered}
N.~C. Jones, R.~Van~Meter, A.~G. Fowler, P.~L. McMahon, J.~Kim, T.~D. Ladd, and Y.~Yamamoto, ``A layered architecture for quantum computing using quantum dots,'' \emph{arXiv preprint arXiv:1010.5022}, 2010.

\bibitem{herrera2010photonic}
D.~A. Herrera-Mart{\'\i}, A.~G. Fowler, D.~Jennings, and T.~Rudolph, ``Photonic implementation for the topological cluster-state quantum computer,'' \emph{Physical Review A}, vol.~82, no.~3, p. 032332, 2010.

\bibitem{divincenzo2009fault}
D.~P. DiVincenzo, ``Fault-tolerant architectures for superconducting qubits,'' \emph{Physica Scripta}, vol. 2009, no. T137, p. 014020, 2009.

\bibitem{groszkowski2011tunable}
P.~Groszkowski, A.~G. Fowler, F.~Motzoi, and F.~K. Wilhelm, ``Tunable coupling between three qubits as a building block for a superconducting quantum computer,'' \emph{Physical Review B}, vol.~84, no.~14, p. 144516, 2011.

\bibitem{kruse2010reconfigurable}
J.~Kruse, C.~Gierl, M.~Schlosser, and G.~Birkl, ``Reconfigurable site-selective manipulation of atomic quantum systems in two-dimensional arrays of dipole traps,'' \emph{Physical Review A}, vol.~81, no.~6, p. 060308, 2010.

\bibitem{yao2012scalable}
N.~Y. Yao, L.~Jiang, A.~V. Gorshkov, P.~C. Maurer, G.~Giedke, J.~I. Cirac, and M.~D. Lukin, ``Scalable architecture for a room temperature solid-state quantum information processor,'' \emph{Nature communications}, vol.~3, no.~1, p. 800, 2012.

\bibitem{kumph2}
M.~Kumph, M.~Brownnutt, and R.~Blatt, ``2 dimensional arrays of rf ion traps with addressable interactions.''

\bibitem{crick2010fast}
D.~Crick, S.~Donnellan, S.~Ananthamurthy, R.~Thompson, and D.~Segal, ``Fast shuttling of ions in a scalable penning trap array,'' \emph{Review of scientific instruments}, vol.~81, no.~1, 2010.

\end{thebibliography}
